\RequirePackage{fix-cm}
\documentclass[twocolumn,epjc3]{svjour3}
\smartqed
\RequirePackage{graphicx}
\RequirePackage{mathptmx}
\RequirePackage{amsmath}
\RequirePackage{amsfonts}
\RequirePackage{amssymb}
\RequirePackage{latexsym}
\RequirePackage{widetext}

\newcommand{\be}{\begin{equation}}
\newcommand{\ee}{\end{equation}}
\newcommand{\bea}{\begin{eqnarray}}
\newcommand{\eea}{\end{eqnarray}}

\DeclareMathOperator\arctanh{arctanh}

\journalname{Eur. Phys. J. C}
\begin{document}

\title{Gauss-Bonnet dark energy on Ho\v{r}ava-Lifshitz cosmology %\thanksref{t1}
}

\author{Samuel Lepe\thanksref{e1,addr1}
        \and
        Giovanni Otalora\thanksref{e2,addr1}}

\thankstext{e1}{e-mail: samuel.lepe@pucv.cl}
\thankstext{e2}{e-mail: giovanni.otalora@pucv.cl}

\institute{Instituto de F\'{\i}sica, Pontificia Universidad Cat\'olica de
Valpara\'{\i}so, Casilla 4950, Valpara\'{\i}so, Chile \label{addr1}}

\date{Received: date / Accepted: date}

\maketitle

\sloppy

\begin{abstract}
We investigate the Gauss-Bonnet dark energy model and its deformed version on Ho\v{r}ava-Lifshitz cosmology, which belongs to the class of cosmologies obtained from the so-called projectable version of Ho\v{r}ava-Lifshitz gravity. In particular, we investigate the bulk/boundary interaction in this scenario through the $Q$ function, which we interpret as a measure of the energy transference between the bulk and the spacetime boundary. Then we discuss whether the thermal equilibrium will be stable or not, once it is reached, and the validity of the generalized second law. We show that the $Q$ function can exhibit sign changes along the cosmic evolution and the Universe reaches the thermal equilibrium as a transient phenomenon.

%\keywords{Gauss-Bonnet invariant \and dark energy \and holographic principle \and Ho\v{r}ava-Lifshitz cosmology}
%\PACS{04.50.Kd, 98.80.-k, 95.36.+x}

\end{abstract}

%%%%%%%%%%%%%%%%%%%%%%%%%%%%%%%%%%%%%%%%%%%%%%%%%%%%
\section{Introduction}\label{Introduction}
%%%%%%%%%%%%%%%%%%%%%%%%%%%%%%%%%%%%%%%%%%%%%%%%%%%%

It is well known that currently the Universe is dominated by dark energy, which provides sufficient negative pressure to accelerate the expansion \cite{Riess:1998cb,Perlmutter:1998np,Ade:2013zuv,Zwicky:1933gu}. The simplest explanation of dark energy is provided by the introduction of a cosmological constant in the Einstein equations, but, this scenario is plagued by a severe fine tuning problem associated with its energy scale \cite{Copeland:2006wr}.  Furthermore, from observational data, it is not ruled out the possibility of a dynamical nature of dark energy with time dependent equation of state (EoS) parameter, which can constitute an interesting alternative to the case cosmological constant, but without a fine tuning problem. Thus, leaving aside the cosmological constant term, there are two principal ways to address a possible dynamical nature of dark energy; it can be interpreted as coming from a modification to General Relativity (GR) by introducing higher curvature terms in the Einstein-Hilbert action (e.g. $f(R)$ theories) \cite{FR-reviews1,FR-reviews2,FR-reviews3,FR-reviews4,FR-reviews5,FR-reviews6,FR-reviews7,Nojiri:2017ncd}, or it can be considered as an exotic matter fluid violating the strong energy condition (e.g. scalar field models) \cite{Copeland:2006wr,AmendolaTsujikawa,Frieman:2008sn1,Frieman:2008sn2,Frieman:2008sn3,Frieman:2008sn4}. Since dark energy currently dominates the cosmic evolution, the investigation of its nature and cosmological properties is essential to our understating of the dynamics of the Universe.  Moreover, if the Universe can be considered a thermodynamic system, being the Hubble Horizon its spacetime boundary, the dominant dark energy component and its interaction with this boundary are fundamental pieces in determining the thermodynamic properties of the Universe and the conditions for its thermal equilibrium. 

A framework which is very useful in studying the bulk/boundary interaction and thermodynamics of the Universe is the so-called Ho\v{r}ava-Lifshitz cosmology \cite{Kiritsis:2009sh,Horava-Lifshitz_Cosmology1,Horava-Lifshitz_Cosmology2,Horava-Lifshitz_Cosmology3}. This class of cosmology is obtained from the framework of Ho\v{r}ava-Lifshitz gravity which is, by abandoning the Lorentz symmetry, a power-counting renormalizable theory of gravitation based in the Lifshitz-type anisotropic scaling at high energy \cite{Horava:2009uw,Blas:2009qj,Blas:2010hb}. Although, it can also be directly obtained from Ho\v{r}ava-Lifshitz $F(R)$ gravity proposed in Refs. \cite{Kluson:2010za,Elizalde:2010ep,Carloni:2010nx,Chaichian:2010yi}. This is the projectable version, where the lapse function depends uniquely on time, and that in spite of the shortcomings related to the existence of an extra scalar degree of freedom with pathological behaviour \cite{Blas:2018flu,Wang:2017brl,Weinfurtner:2010hz,Sotiriou:2010wn,Koyama:2009hc,Wang:2009yz}, it constitutes an useful tool to gain insight in the most complete versions of the theory and their consequences for classical and quantum cosmology \cite{Bernardini:2017jlz,Saridakis:2011pk,Bertolami:2011ka,Mukohyama:2009mz,Sotiriou:2009bx,Maeda:2010ke,Wang:2009rw,Brandenberger:2009yt,Calcagni:2009ar,Cruz:2015lsz}. So, unlike GR, the fundamental symmetry is the invariance under foliation preserving diffeomorphism, that is to say, three-dimensional spatial diffeomorphism plus space-independent time reparametrization, which implies three local momentum constraints and one global Hamiltonian constraint integrated over the whole space at each time. Hence, in the Friedmann-Robertson-Walker (FRW) spacetime, the lack of a local Hamiltonian constraint leads to a loss of the first Friedmann equation ($00$ component), and therefore, the two independent dynamical equations are now the second Friedmann equation ($ii$ component) and the nonconservation equation satisfied by the cosmic fluid in our patch of the Universe inside the Hubble horizon. The $Q$ function appearing in the right hand side of this last equation, and which represents the nonconservation of energy, can be physically interpreted as a measure of the interchange of energy between the bulk (observable Universe) and the Hubble horizon \cite{Cruz:2015lsz}. There are two ways in determining the specific functional form of $Q$. One could follow the usually considered approach for the treatment of two interacting fluids in GR, where the functional form of the coupling term is imposed by hand \cite{AmendolaTsujikawa}. Also, a second approach could consist in choosing an ansatz for the energy density of the bulk, through some physical arguments, as for example, those behind the holographic philosophy \cite{Cruz:2015lsz}.

In choosing a candidate for dark energy we have very interesting possibilities. From the viewpoint of the Holographic approach, the dark energy density is proposed to saturate the holographic bound, applied to the size of the current Universe \cite{Li:2004rb,Nojiri:2005pu,Nojiri:2017opc}. For example, in Ref. \cite{Nojiri:2017opc}, it was proposed a generalized holographic dark energy model where the infrared cutoff is identified with the combination of the characteristic FRW spacetime parameters, such as, the Hubble rate, particle and future horizons, cosmological constant, the universe lifetime and their derivatives. However, if one identifies the characteristic length scale as the Hubble scale, it is seen that although this proposal matches the currently observational value of dark energy density, its equation of state (EoS) is out of the observationally allowed range \cite{Hsu:2004ri}. Recently, in Ref. \cite{Granda:2013gka} it has been proposed a dark energy density based on the Gauss-Bonnet four-dimensional invariant and its modification, which can be physically interpreted as an energy density nonsaturating the holographic principle \cite{Bekenstein:1973ur,tHooft:1993dmi,Bousso:1999xy,Cohen:1998zx,Susskind:1994vu}, and it is also a very special variant of the most general infrared cutoff established in Ref. \cite{Nojiri:2017opc}. The quadratic curvature Gauss-Bonnet term is a topological invariant which is strongly motivated from higher derivative gravity theories \cite{Stelle:1976gc}, being that it appears in QFT renormalization in curved spacetimes \cite{Birrell:1982ix,Armaleo:2017lgr}, and it is one of the most promising candidates to provide the leading-order correction to the low energy string gravity \cite{Gross:1986mw,Barrau:2003tk}. Furthermore, as it was shown in Ref. \cite{Granda:2013gka}, this Gauss-Bonnet cutoff allows us to obtain the current values of dark energy density and a dark energy EoS parameter compatible with the observational data. 

In the present paper we study the thermodynamics of the Gauss-Bonnet (GB) dark energy model and its deformed version in the framework of Ho\v{r}ava-Lifshitz cosmology. We study the bulk-boundary interaction, paying particular attention in the stability of thermal equilibrium and the generalized second law. Throughout the paper, we adopt units $8 \pi G=1=c$.

\section{Gauss-Bonnet dark energy on flat Ho\v{r}ava-Lifshitz cosmology}

\subsection{The $Q$ function}

In Ho\v{r}ava-Lifshitz cosmology \cite{Kiritsis:2009sh,Horava-Lifshitz_Cosmology1,Horava-Lifshitz_Cosmology2,Horava-Lifshitz_Cosmology3} the field equations for the cosmic fluid, with barotropic equation of state $p=\omega \rho$, are given by
\begin{eqnarray}
\eta \left( 3H^{2}+2\dot{H}\right) &=&-\omega \rho,\label{hl1} \\
\dot{\rho}+3H\left( 1+\omega \right) \rho &=&-Q,  \label{hl2}
\end{eqnarray}%
where $\eta$ is a dimensionless constant parameter associated to 
diffeomorphism invariance which is confined to the range $0<\eta <1$ in the case of ghost graviton, $\eta <0$ or $\eta >1$ for no-ghost graviton, and $\eta$ is fixed to $1$ in GR \cite{Horava:2009uw}. The square speed of the scalar graviton is $c^2_{s}=(1-\eta)/3\eta$, and from observational constraints and stability considerations one obtains extremely tight bounds on the parameters of projectable Ho\v{r}ava-Lifshitz gravity, such that, $|\eta-1|<10^{-60}$ or equivalently $\left|c_{s}\right|<10^{-30} $ \cite{Clifton:2011jh}. However, in spite of they are extremely small numbers, it is worth remembering that the better ``zero'' in Physics is $\Lambda l_{p}^2\sim 10^{-120}$, being $\Lambda$ the cosmological constant and $l_{p}$ the Planck length \cite{Martin:2012bt}.

From Eqs. \eqref{hl1} and \eqref{hl2} we find 
\be
\frac{Q}{3\eta H^{3}}=\frac{2}{3\omega }\left( \frac{3}{4\pi }A_{h}\dot{H}+\frac{\ddot{H}}{H^{3}}\right) -\frac{A_{h}\left( 1+\omega \right) \rho }{4\pi \eta }, 
\label{Q}
\ee where $A_{h}=4\pi H^{-2}$ is the area of Hubble horizon. 
This $Q$ function, which represents the rate of energy nonconservation, can be physically interpreted as a measure of the flux of energy between the bulk and the horizon \cite{Cruz:2015lsz,Lepe:2015ila,Mimoso:2016jwg}. In the bulk exists one only component characterized by the EoS parameter $w$. % The subindex ``$b$'' makes mention of that one component present in the bulk, interpreted as the dominant component. 
Since at late times the evolution of the background cosmology is dominated not by matter, but rather by vacuum energy or dark energy, here we limit ourselves to the case $w=-1$. The low energy limit can be recovered only if $Q\rightarrow 0$ \cite{Horava-Lifshitz_Cosmology1}. Also, it becomes a fundamental ingredient in the time evolution of the cosmological parameters and the dynamics of the Universe. For example, by using Eqs. \eqref{hl1} and \eqref{hl2}, we find that the effective EoS parameter of the cosmic fluid is given by 
\be
w_{eff}=w\left[1+\frac{1}{2\left(q-\frac{1}{2}\right)}\left(\frac{Q}{3 \eta H^3}\right)\right],
\label{weff}
\ee where $q=-1-\dot{H}/H^2$ is the deceleration parameter, and the 
$Q$ function appears as a determinant factor in the dynamical behaviour of this cosmological parameter. 

\subsection{Gauss-Bonnet cutoff}

There are two ways in determining the $Q$ function. The first way, it is that usually considered in GR and coupled dark energy theories, where the specific form of the coupling function $Q$ between scalar field and matter is introduced by hand, as for example, through the ansatz $Q\sim \rho \dot{\phi}$ with $\dot{\phi}$ the velocity of the homogeneous scalar field \cite{Copeland:2006wr,Amendola:1999er,Gumjudpai:2005ry} or $Q\sim \Gamma \rho$ with the normalization of the factor $\Gamma$ in terms of the Hubble parameter $H$, i.e. $\Gamma/H =\eta$ , where $\eta$ is a dimensionless constant \cite{AmendolaTsujikawa}. On the other hand, a second approach consists in determining the $Q$ function
through Eq. \eqref{hl2}, or equivalently, via Eq. \eqref{Q}, by considering a phenomenological ansatz for the dark energy density $\rho$, as for example, that one obtained from the holographic philosophy \cite{Cruz:2015lsz}.

It is well known that for an effective quantum field theory in a box of size
$L$ with ultraviolet (UV) cutoff $\Lambda$ the entropy $S$ scales extensively with the volume as $S\sim L^3 \Lambda^3$ \cite{Susskind:1994vu}. On the other hand, the holographic principle as given by the Bekenstein bound \cite{Bekenstein:1973ur}, postulates that the maximum entropy in a box of volume $L^3$ behaves nonextensively and proportional to the area of the box. In order to reconcile quantum field theory with the holographic principle, in Ref. \cite{Cohen:1998zx} the authors have imposed a relationship between UV and infrared (IR) cutoffs. This relationship was established by putting a stronger constraint on IR cutoff $1/L$, in such a form that if $\rho$ is the quantum zero-point energy density due to the UV cutoff $\Lambda$, the total energy in a region of size $L$ should not exceed the mass of a black hole of the same size, that is, $L^3 \rho \leq L M_{p}^2$, or equivalently $\rho \leq L^{-2} M_{p}^2$.

If one associates $L$ with the size of the current Universe, then the vacuum energy $\rho= L^{-2} M_{p}^2 $ related to this holographic principle, may be identified as dark energy, or, also usually the so-called holographic dark energy density. However, identifying the Hubble scale $1/H$ as the characteristic length scale $L$, one can see that although this proposal matches the currently observational value of dark energy density, $\rho= M_{p}^2 H_{0}^2$, its EoS parameter is out of the observationally allowed range \cite{Hsu:2004ri}. So, in Ref. \cite{Granda:2013gka}, the author has proposed a novel cutoff which is not based in the Hubble scale, but whether in the Gauss-Bonnet (GB) four-dimensional invariant and its modification. The quadratic curvature Gauss-Bonnet term is a topological invariant which surges in higher derivative gravity theories, where it has been shown that gravitational actions which include terms quadratic in the curvature tensor are renormalizable \cite{Stelle:1976gc}. Thus, following the holographic philosophy, the proposal in Ref. \cite{Granda:2013gka} postulates the relation $\rho =\alpha\,GB$ where $GB=R^{2}-4R_{\mu \nu }R^{\mu \nu }+R_{\mu \nu \lambda \rho }R^{\mu\nu \lambda \rho }=24\left( H^{2}+\dot{H}\right) H^{2}$  is the GB invariant, and for the "deformation" $\rho =24\left( \alpha H^{2}+\beta \dot{H}\right) H^{2}$. 
In this way, this GB cutoff provides a holographic dark energy density which scales in a natural form as $H^{4}$ and not as $H^2$, and hence, it may be physically interpreted as a dark energy density nonsaturating the holographic principle once that one always has the relation $M_{p}^2\gg H^2$. Furthermore, as it was shown in Ref. \cite{Granda:2013gka}, this GB dark energy density explains the current values of dark energy, and at the same time, its dark energy EoS parameter is compatible with the observational constraints.

Below, we study these holographic candidates to dark energy in the context of Ho\v{r}ava-Lifshitz cosmology, paying particular attention in the $Q$ function and its behaviour along the cosmic evolution.

\subsubsection{Gauss-Bonnet dark energy}

We discuss the dark energy cutoff given by \cite{Granda:2013gka}

\begin{equation}
\rho =24\lambda H_{0}^{-2}\left( H^{2}+\dot{H}\right) H^{2}=-24\lambda
H_{0}^{-2}qH^{4}.  
\label{hl3}
\end{equation} By replacing \eqref{hl3} in
\eqref{hl1}, we can write
\begin{equation}
\frac{1/2-q}{q}=\frac{1}{2} A H^{2}, 
\label{hl4}
\end{equation}%
where $A=24\lambda H_{0}^{-2}\omega /\eta $. But $\rho >0$ implies
$q<0$ so that we must have 

\begin{equation}
A<0\Longrightarrow \left\{ 
\begin{array}{c}
\omega >0\text{ and }\eta <0,\text{ or} \\ 
\omega <0\text{ and }\eta >0.%
\end{array}%
\right.
\label{hl5}
\end{equation}%

So, according to \eqref{hl1} and \eqref{hl3}, we can write
\be
\dot{H}=-\left( \frac{3-\left\vert A\right\vert H^{2}}{2-\left\vert
A\right\vert H^{2}}\right) H^{2}, 
\label{dfEq}
\ee
where the implicit solution for $E(t)=H\left( t\right) /H_{0}$ is given by
\begin{eqnarray}
&&\frac{1}{E\left( t\right) }+\sqrt{2\frac{\lambda }{\eta }\left\vert
\omega \right\vert}\Big[ \arctanh \left( \sqrt{8\frac{\lambda }{\eta }%
\left\vert \omega \right\vert }E\left( t\right) \right) \nonumber\\
&&-\arctanh
\left( \sqrt{8\frac{\lambda }{\eta }\left\vert \omega \right\vert }\right) %
\Big] =1+\frac{3}{2}H_{0}t_{0}\left( \frac{t}{t_{0}}-1\right) . 
\label{hl6}
\end{eqnarray} From this equation we can study numerically the late-times behaviour of the Hubble function $H(t)$.  
Moreover, by substituting Eq. \eqref{hl3} into Eq. \eqref{hl2}, we obtain the coupling function $Q$ as 
\bea
&& \frac{Q}{3 \eta H^3}=\frac{8 \lambda}{\eta} \Bigg[E\left(3 (w+1) q E+\frac{\dot{q}}{H_{0}}\right)+ \frac{4 q\dot{E}}{H_{0}}\Bigg],
\label{QGB}
\eea and using the numerical solution for $H(t)$ from Eq. \eqref{hl6} we obtain $Q(t)$.

\begin{figure}[htbp]
\begin{center}
\includegraphics[width=0.45\textwidth]{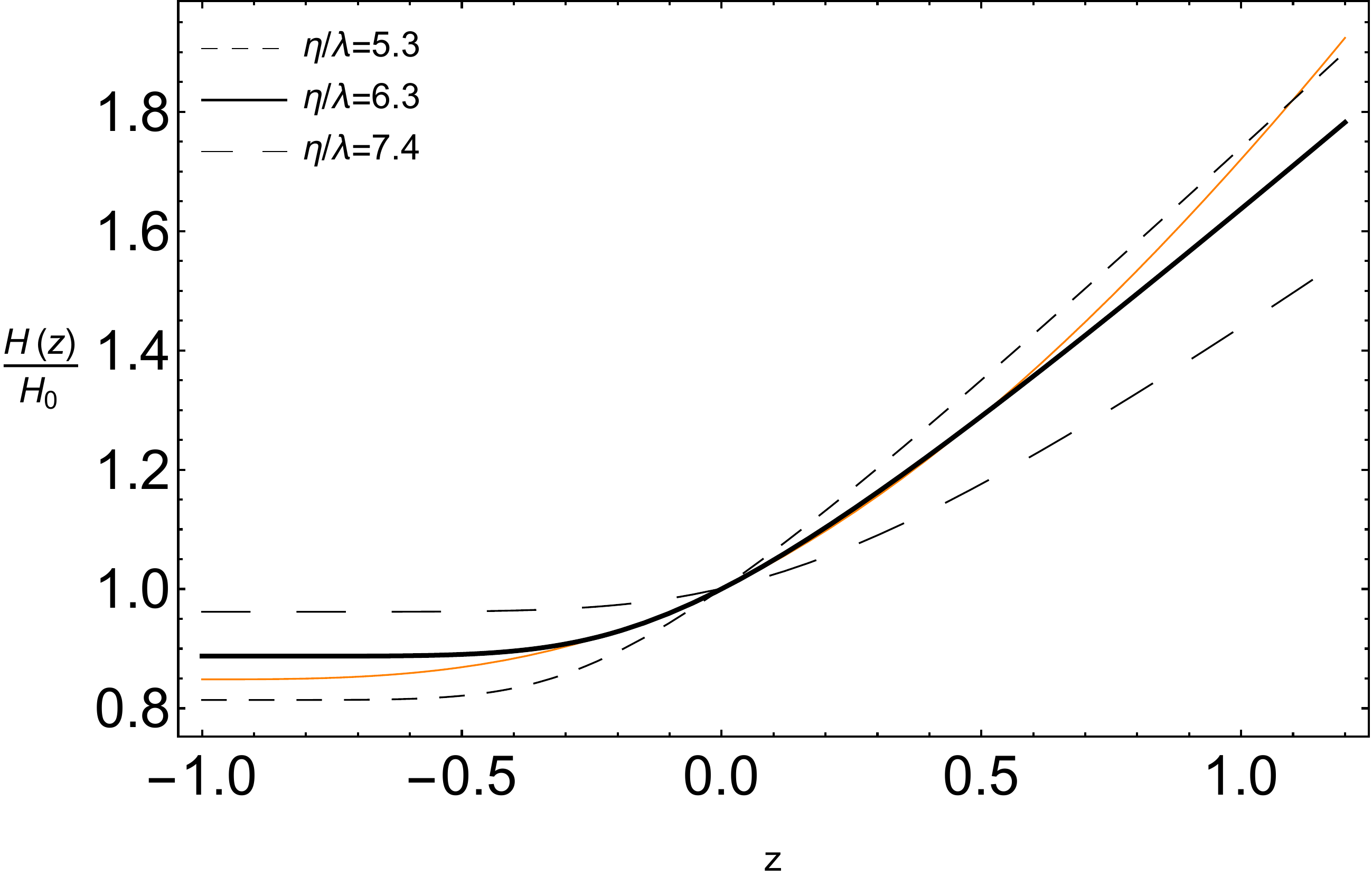}
\includegraphics[width=0.45\textwidth]{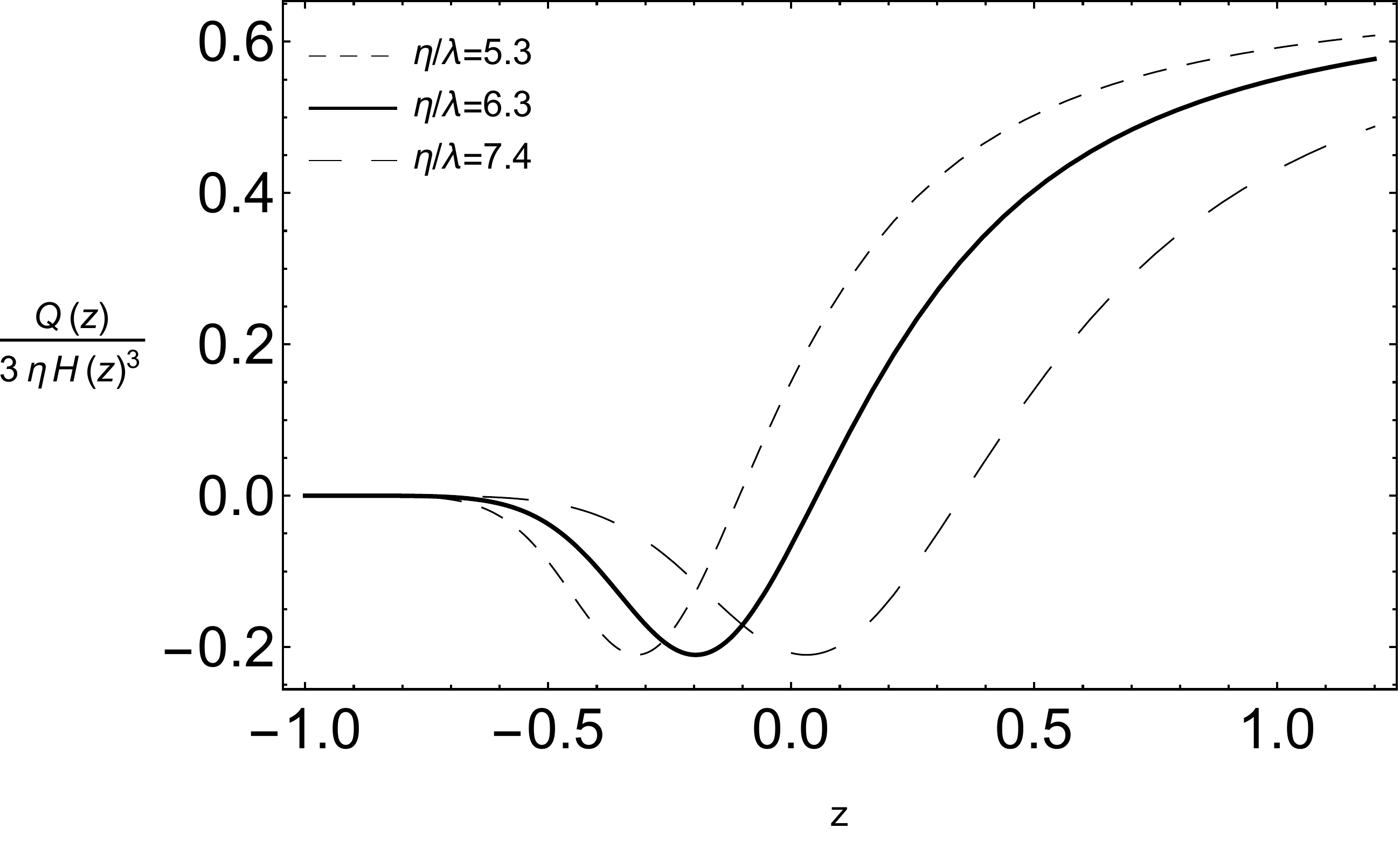}
\caption{\it{For the GB DE model it is shown the evolution of the Hubble parameter $H(z)$ (upper graph) and  the coupling $Q(z)$ (lower graph) as functions of the redshift $z$ for the fixed value $w=-1$ and several different values of the parameters $\eta$ and $\lambda$. From Eq. \eqref{ConstGBDE}, for $\eta/\lambda=\{5.3, 6.3, 7.4\}$ we obtain $q(0)\approx\{-0.40,-0.55,-0.80\}$. The orange line corresponds to the evolution of the $\Lambda$CDM model.}}
\label{FIG1}
\end{center}
\end{figure}

From Eq. \eqref{hl4} we can write
\begin{equation}
\frac{\eta }{\lambda }=12\left\vert \omega \right\vert
\left( \frac{\left\vert q\left( 0\right) \right\vert }{\left\vert q\left(
0\right) \right\vert +1/2}\right), 
\label{ConstGBDE}
\end{equation}%
and then we must have $\eta >0$. So, according to the observational data,
something we could say about the quotient $\eta /\lambda $. For instance, by using $q(0)\approx -0.8$ and $\omega= -1$, we have $%
\eta /\lambda \approx 7.4$ and $\lambda $ could decide whether there is or not ghost. We also can introduce the redshift $z=a_{0}/a(t)-1$ as independent variable instead of cosmic time $t$, and then apply the transformation $d/dt=-(1+z)H(z)d/dz$, in order to obtain $H(z)$ and $Q(z)$. In FIG \ref{FIG1} it is shown the evolution of $H(z)$ (upper graph) and $Q(z)$ (lower graph) for some values of the parameters $w=-1$ and
$\eta/\lambda=\{5.3, 6.3, 7.4\}$, being that in this case one obtains $q(0)\approx\{-0.40,-0.55,-0.80\}$. Also, for $\eta \sim 1 $ one can see that $\lambda\approx \{0.19, 0.16, 0.14\}$ and therefore it satisfies the constraint $\lambda<1$.
It is observed a good concordance with $\Lambda$CDM scenario at low redshifts ($z<1.2$) once that dark energy becomes the dominant component. In the case of GB dark energy model, the most favoured values by the observational data correspond to the values $\eta/\lambda=6.3$, being $w=-1$ and $q(0)=-0.55$. We see that the $Q$ function exhibits sign changes along the cosmic evolution, taking negative values in the present epoch, at $z=0$. Additionally, from Eq. \eqref{weff}, and by using the values of the quotient $Q(z)/(3\eta H(z))$, we can evaluate the effective EoS parameter as a function of the redshift $z$. In particular, we see that it matches the observational constraints for the EoS parameter, taking a phantom value $w_{eff}(0)\approx -1.03$ at $z=0$.

\subsubsection{Deformed Gauss-Bonnet dark energy}

We discuss now the ``deformation'' of \eqref{hl3} given by \cite{Granda:2013gka}
\begin{equation}
\rho =24\left( \alpha H^{2}+\beta \dot{H}\right) H^{2}=24H_{0}^{-2}\left[ 
\bar{\alpha}-\bar{\beta}\left( 1+q\right) \right] H^{4}, 
\label{hl12}
\end{equation}%
and $\rho >0\Longrightarrow \bar{\alpha}>\bar{\beta}\left( 1+q\right) $, $%
\forall q$. Here, we have rescaled the parameters in the form $\alpha=\bar{\alpha}/H_{0}^2$ and $\beta=\bar{\beta}/H_{0}^2$. By using \eqref{hl12} in \eqref{hl1}, we have

\begin{equation}
\frac{1}{12}\left( \frac{1/2-q}{\bar{\alpha}-\bar{\beta}\left( 1+q\right) }%
\right) =-\left( \frac{\omega }{\eta }\right) \left( \frac{H}{H_{0}}\right)
^{2},  
\label{hl13}
\end{equation}%
and then $sgn\left( 1/2-q\right) =-sgn\left( \omega /\eta \right) $, i. e.,

\begin{eqnarray}
q &>&1/2\Longrightarrow sgn\left( \omega /\eta \right) =1,  \label{hl14}
\\
q &<&1/2\Longrightarrow sgn\left( \omega /\eta \right) =-1.  \label{hl15}
\end{eqnarray}%
By seeing \eqref{hl13}, today
\begin{equation}
\eta =12\left\vert \omega \right\vert \bar{\beta}\left( 
\frac{\bar{\alpha}/\bar{\beta}-\left( 1-\left\vert q\left( 0\right)
\right\vert \right) }{1/2+\left\vert q\left( 0\right) \right\vert }\right)
\Longrightarrow \eta >0,  
\label{hl17}
\end{equation}
and we have a somewhat more complicated situation than \eqref{ConstGBDE} in order
to visualize values for both $\bar{\alpha}$ and $\bar{\beta}$ from the
observational data and then values for $\eta $. We recall that $0<\eta <1$
(ghost graviton) or $\eta >1$ (no-ghost graviton) and $\eta$ is fixed to $1$ in GR. Even more, according to \eqref{hl15} and \eqref{hl17}, $\eta >0\Longrightarrow \omega <0$. Here, we can not forget that $
0<\left\vert q\left( 0\right) \right\vert <1$. From the inequality $\bar{\alpha}>\bar{\beta}\left( 1+q\right)
\rightarrow q<-1+\bar{\alpha}/\bar{\beta}$ we could have, $q<0$ $\left( \bar{%
\alpha}/\bar{\beta}<1\right) $ and $q>0$ $\left( \bar{\alpha}/\bar{\beta}%
>1\right) $. We note that, unlike \eqref{hl3}, $q<0$, we have now the possibility of $q>0$. By taking $q(0)\simeq-0.55$, $w=-1$ in Eq. \eqref{hl17} we obtain $\eta/\bar{\beta}\approx 11.43 \left(\bar{\alpha}/\bar{\beta}-0.45\right)$. For $\eta\sim 1$, and $\bar{\alpha}/\bar{\beta}=\{0.85, 0.65, 0.55\}$ we obtain $\bar{\beta}\approx \{0.22, 0.44,0.88\}$, respectively.

According to \eqref{hl1} and \eqref{hl12}, we have

\begin{equation}
\dot{H}=-\left( \frac{3+A\bar{\alpha}H^{2}}{2+A\bar{\beta}H^{2}}\right) H^{2}
\label{hl18}
\end{equation}%
where $A=24 H_{0}^{-2}\left( \omega /\eta \right) $ and its sign depends on the expression $sgn\left( \omega /\eta \right) $. According to \eqref{hl14}, $A>0$ if $q>1/2$
and $A<0$ if $q<1/2$. So, from \eqref{hl18} with $A>0$, we obtain

\begin{equation}
\frac{1}{H}+\frac{3A}{2}\left( \frac{2\bar{\alpha}}{3}-\bar{\beta}\right)
\int \frac{dH}{3+A\bar{\alpha}H^{2}}=\frac{3}{2}t+const., 
\label{hl19}
\end{equation}%
and from here $H\left( t\right) $.

\begin{figure}[htbp]
\begin{center}
\includegraphics[width=0.45\textwidth]{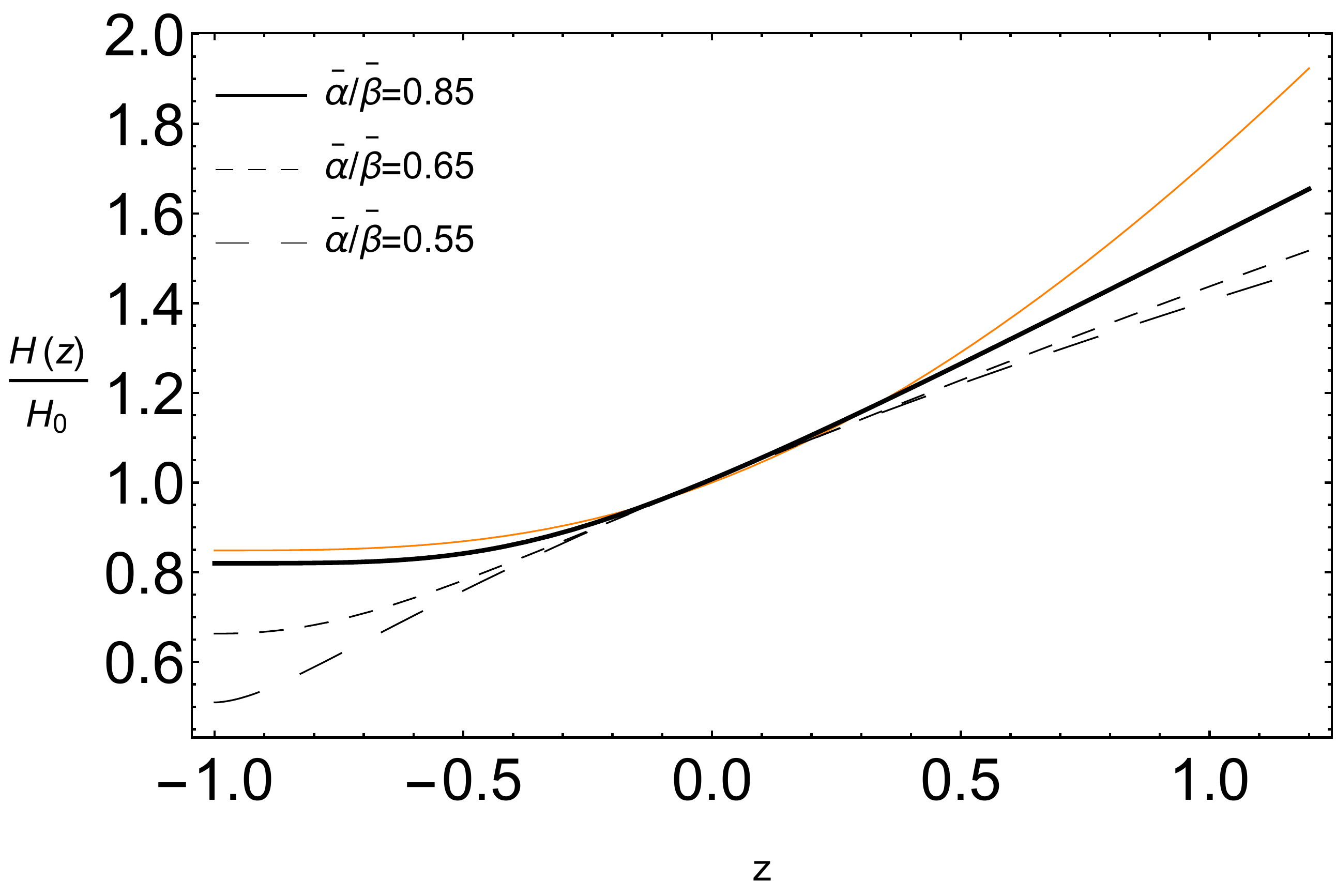}
\includegraphics[width=0.45\textwidth]{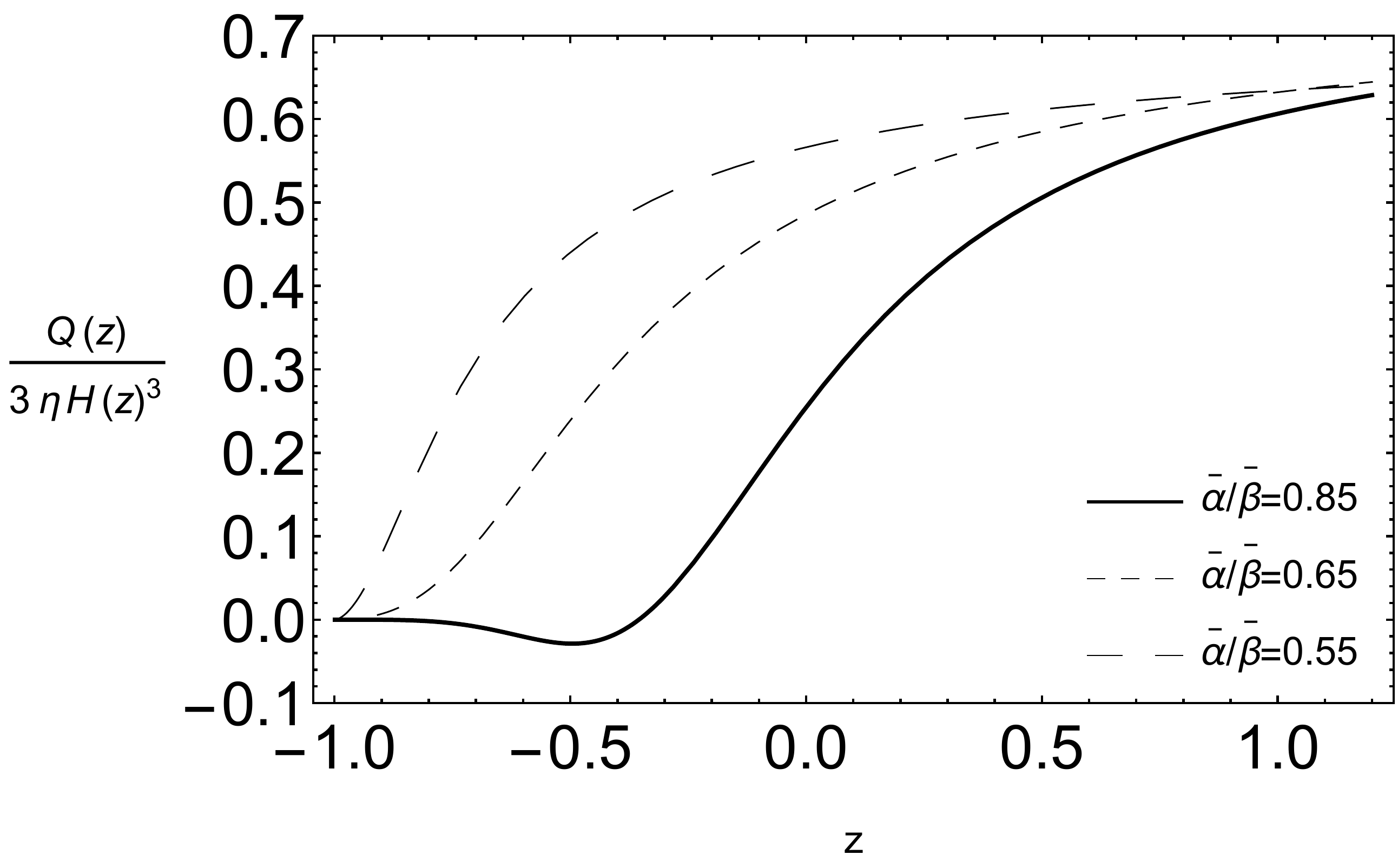}
\caption{\it{For the deformed GB DE model it is shown the evolution of the Hubble parameter $H(z)$ (upper graph) and  the coupling $Q(z)$ (lower graph) as functions of the redshift $z$ for the fixed values $w=-1$ , $\eta\sim 1$, $q(0)=-0.55$ and several different values of the quotient $\bar{\alpha}/\bar{\beta}$. The orange line corresponds to the evolution of the $\Lambda$CDM model.}}
\label{FIG3}
\end{center}
\end{figure}

If $A<0$, we obtain
\begin{equation}
\frac{1}{H}-\frac{3\left\vert A\right\vert }{2}\left( \frac{2}{3}\bar{\alpha}%
-\bar{\beta}\right) \int \frac{dH}{3-\left\vert A\right\vert \bar{\alpha}%
H^{2}}=\frac{3}{2}t+const.,
\label{hl20}
\end{equation}%
and from here $H\left( t\right)$. So, from $H\left( t\right) $ we obtain $Q\left( t\right)$ by substituting \eqref{hl12} in Eq. \eqref{hl2}, and it reads as
\bea
&&\frac{Q}{3\eta H^3}=\frac{8}{\eta} \Bigg[\frac{\bar{\beta}  E \dot{q}}{H_{0}}+\left(\bar{\beta}\left(1+q\right)-\bar{\alpha}\right)\times \nonumber\\
&& \left(3 (w+1) E^2 +\frac{4 \dot{E}}{H_{0}}\right)\Bigg].
\label{QMDE}
\eea Since we are interested in the dark-energy dominated era, we take Eq. \eqref{hl20} as the evolution equation for $H$ for $A<0$ ($w<0$ and $\eta>0$). In FIG \ref{FIG3} (upper graph) we depict the evolution of $H(z)$ as a function of $z$. Also, in FIG \ref{FIG3} (lower graph) it is shown the cosmic evolution of $Q(z)$ in according to Eq. \eqref{QMDE}. It is shown that the deformed GB model can fit quite well the $\Lambda$CDM scenario for small deviations from the nondeformed case, that is to say, for $\bar{\alpha}/\bar{\beta}\lesssim 1$,  at low redshifts ($z<1.2$), once that dark energy becomes the dominant component. As in the nondeformed case, the $Q(z)$ function presents sign changes along the cosmic evolution. However, in this  case, $Q(z)$ takes positive values in the present, at $z=0$, for the most favoured values of the parameters $\bar{\alpha}/\bar{\beta}=0.85$. 
This result is important because positive values of $Q(z)$ are also favoured from observational data, see e.g. Refs. \cite{Costa:2016tpb,Wang:2016lxa}. 

\section{Thermal equilibrium and Second law}\label{Thermo}

In the context of Ho\v{r}ava-Lifshitz cosmology, a $Q$ function is naturally present into the theory. Thus, the Universe can be seen as a thermodynamic system composed of the two mutually interacting parts, the bulk and the boundary. In order to study the classical thermodynamics of this composed system, we must introduce the temperature and entropy of each one of the parts. The temperature and entropy of the bulk are calculated using the Gibbs equation and the integrability condition applied to the Hubble volume enclosing the cosmic fluid with effective EoS parameter $w_{eff}$ given by Eq. \eqref{weff}. In this case, the equation for the entropy reads as \cite{Cruz:2015lsz}
\bea
&&T_{b}\dot{S}_{b}=\frac{8 \pi}{3}\eta \left(\frac{q-1/2}{\omega}\right)\times \nonumber\\
&&\left[3\left(1+\omega\right)q-\frac{3 \omega}{2\left(q-1/2\right)}\left(\frac{Q}{3 \eta H^3}\right)\right],
\label{Sb}
\eea whereas that the temperature is given by 
\bea
&& \frac{T_{b}(z)}{C_{0}(1+z)^{3 w}}=\exp{\left[\int{\frac{3 w}{2\left(q-\frac{1}{2}\right)}\left(\frac{Q}{3\eta H^3}\right)\frac{dz}{1+z}}\right]},
\label{Tb}
\eea  where $C_{0}$ is an integration constant. On the other hand, the temperature and entropy of the boundary, is calculated using the Holographic principle which establishes that the temperature of the Horizon is $T_{h}=H/(2 \pi)$ and its associated entropy is $S_{h}=8 \pi^2 H^{-2}$ \cite{Cai:2005ra1,Cai:2005ra2,Cai:2005ra3,Cai:2005ra4}.

A first point to be studied is the question of whether a stable thermal equilibrium between the bulk and the horizon can be reached or not. In order to discuss the stability of thermal equilibrium we introduce the heat capacity of the bulk, $C_{b}=T_{b}\left(\partial S_{b}/\partial T_{b}\right)$, and the heat capacity of the boundary, $C_{h}=T_{h}\left(\partial S_{h}/\partial T_{h}\right)$. From Eq. \eqref{Sb} we obtain the heat capacity of the bulk which is written as
\be
C_{b}=-\left(\frac{1}{6 \pi T_{b} T_{h}\omega_{eff}}\right)f(z),
\label{Cb}
\ee where $f(z)$ is defined by the rhs of Eq. \eqref{Sb}, which depends also on the coupling function $Q$. Moreover, for the heat capacity of the boundary we find $C_{h}=-4/T_{h}^2<0$. 

The heat capacity of the boundary is always negative, whereas that the heat capacity of bulk does not have a definite sign and it depends on the sign of the function $f(z)$, that is, it depends on the cosmic evolution of the $Q$ function. Since we are dealing with systems that present negative heat capacities, as for example the horizon, or perhaps also the bulk, let us remember when the stable thermal equilibrium can be attained for theses systems in thermal contact. Clearly, there are mainly two criteria that we must consider in the moment \cite{LyndenBell:1998fr,LVelazquez}: 
\begin{enumerate}
	\item Two systems with negative heat capacities in thermal contact do not attain thermal equilibrium.
	\item A system of negative heat capacity can achieve a stable thermal equilibrium in contact with one other of positive heat capacity if their combined heat capacity is negative.
\end{enumerate}
The second point to be verified is the fulfilment of generalized second law. It determines that the total entropy of the system, the entropy of the bulk $S_{b}$ plus the entropy of the boundary $S_{h}$, never decreases, that is, we always must have $\dot{S}_{h}+\dot{S}_{b}\geq 0$. Thus, by using the equations \eqref{Sb} and \eqref{Tb}, along with the relation $T_{h}\dot{S}_{h}=8 \pi (1+q)$ for the Horizon, the generalized second law puts the following constraint on the $Q$ function
\bea
R=\left(\frac{Q}{3 \eta H^3}\right)-D\leq 0,
\label{R}
\eea with
\be
D=2 \Bigg[\left(\frac{q-1/2}{\omega}\right)
q \left(1+\omega\right)+\left(\frac{1+q}{\eta}\right)\left(\frac{T_{b}}{T_{h}}\right)\Bigg],
\ee where $T_{b}$ is given by \eqref{Tb} and $T_{h}=H/(2 \pi)$.
Here we also have used the fact that $\rho H^{-1}/(2\eta)=(q-1/2)/\omega>0$, as it can be see from Eq. \eqref{hl2}.

Below, following the same scheme of section II, we analyse separately the cases of GB model and its deformed version.

\subsection{Gauss-Bonnet dark energy}

\begin{figure}[htbp]
\begin{center}
\includegraphics[width=0.45\textwidth]{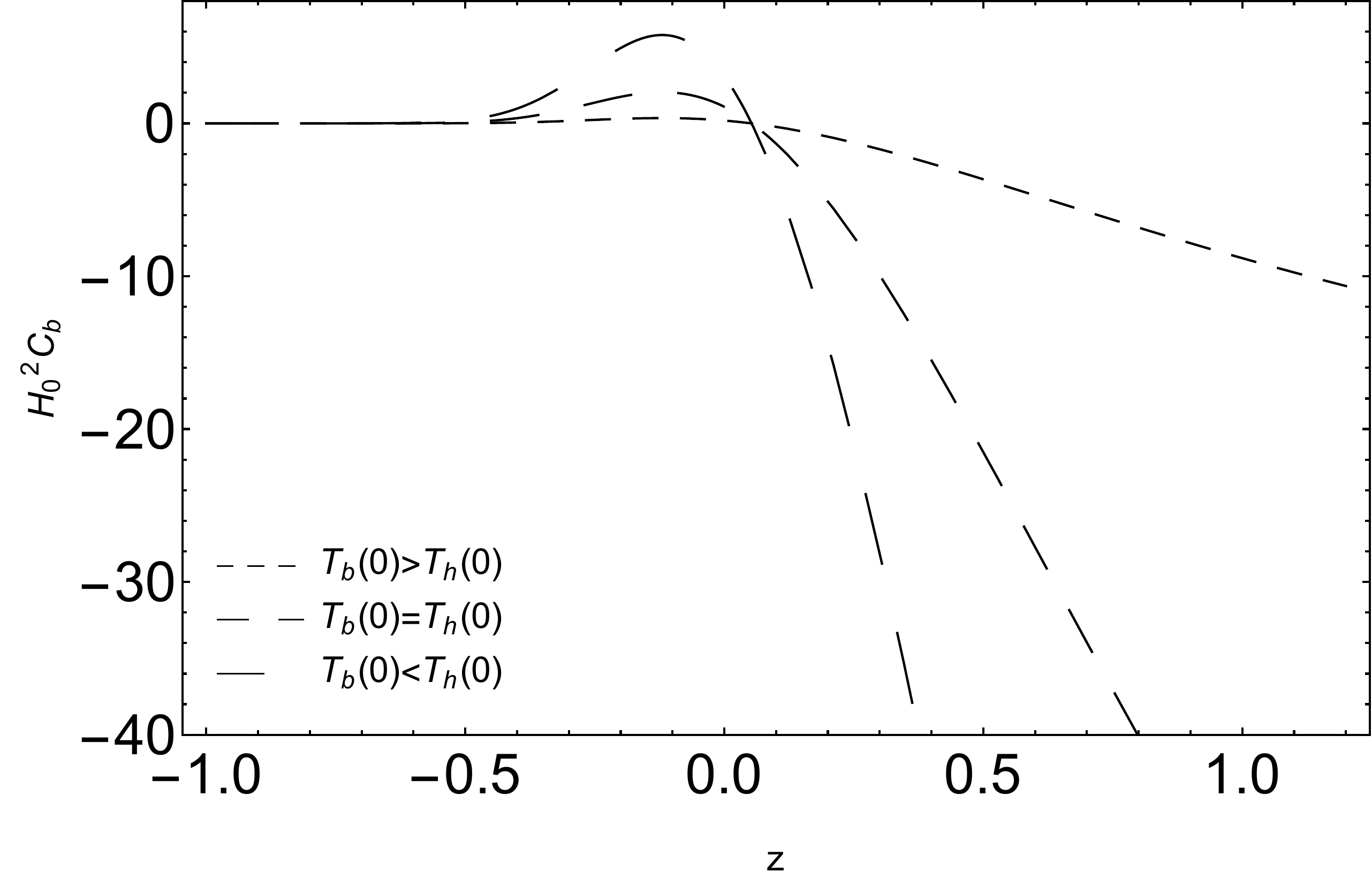}
\includegraphics[width=0.45\textwidth]{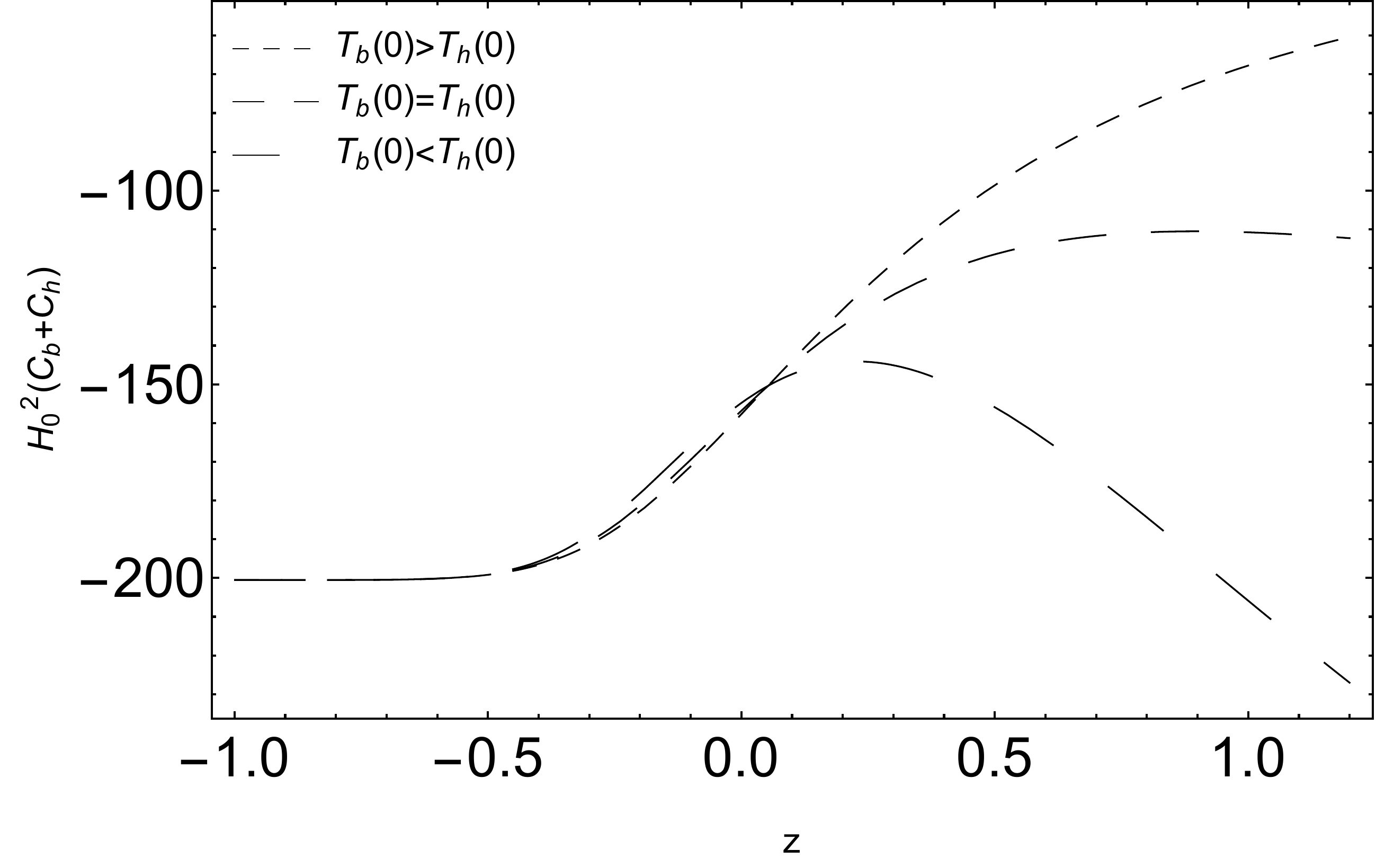}
\caption{\it{GB Model: In the upper graph we show the evolution of the heat capacity of the bulk. On the other hand, in the lower graph we depict the evolution of the sum of capacities of the bulk and the horizon. Following the criteria $1$ and $2$ exposed in section \ref{Thermo}, we observe that the thermal equilibrium is possible as a transient phenomenon.}}
\label{FIG4}
\end{center}
\end{figure}

As seen above, in order to discuss the stable thermal equilibrium of the system bulk/boundary, it is necessary to identify the sign of the heat capacity of the bulk along the cosmic evolution. It does not have a definite sign because it depends on the cosmic evolution of the $Q$ function. In FIG \ref{FIG4} (upper graph) we depict the cosmic evolution of the heat capacity of the bulk. We see that it takes negative values until short time before the present epoch $z=0$, for after to go through a sign change and then to take positive values during a transient epoch, which extends into the future. On the other hand, also in FIG \ref{FIG4} (lower graph), we depict the cosmic evolution of the sum of heat capacities of the bulk and the boundary, observing that it takes negatives values in all the redshifts. Therefore, following the criteria exposed in section \ref{Thermo}, we can observe that in the case of GB dark energy model, the thermal equilibrium can be reached only as a transient phenomenon, once that $C_{b}$ takes positive values during some time and the sum of capacities $C_{b}+C_{h}$ is negative. This happens independently of the set of initial conditions $T_{b}(0)>T_{h}(0)$, $T_{b}(0)=T_{h}(0)$ or $T_{b}(0)<T_{h}(0)$, for the temperature of the bulk and the boundary.

\begin{figure}[htbp]
\includegraphics[width=0.45\textwidth]{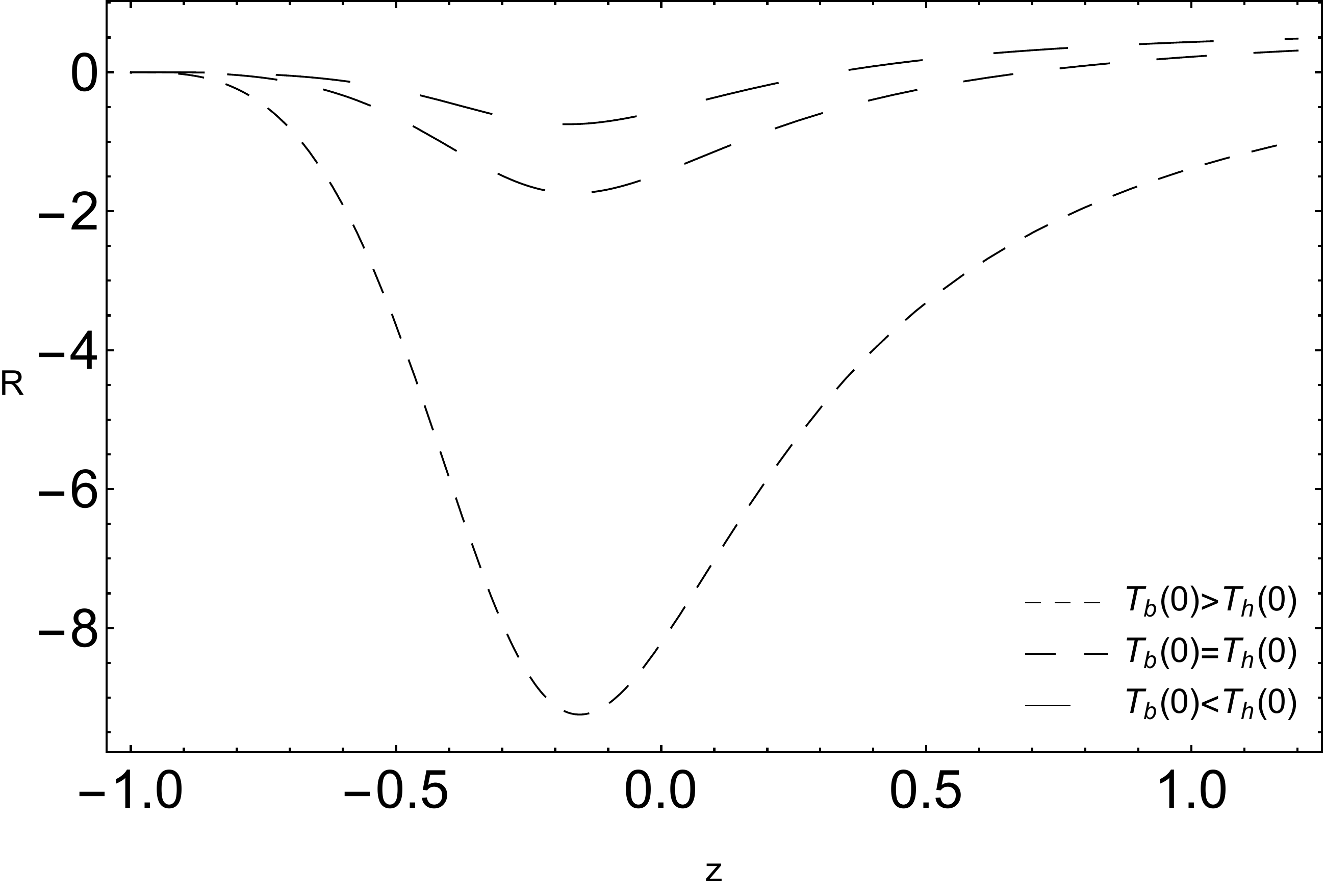}
\caption{\it{GB Model: Evolution of parameter $R$ defined in Eq. \eqref{R}, which allows us to express the nonviolation of the generalized second law through the constraint $R\leq 0$. We observe that there is a violation of the generalized second law at higher redshifts for the initial conditions $T_{b}(0)=T_{h}(0)$ or $T_{b}(0)<T_{h}(0)$, while for $T_{b}(0)>T_{h}(0)$ apparently there is not violation.}}
\label{FIG5}
\end{figure}

In relation to the fulfilment of generalized second law, in FIG \ref{FIG5} we depict the behaviour of the parameter $R$ defined in \eqref{R}. At higher redshifts we observe a violation of the second law for the initial conditions $T_{b}(0)=T_{h}(0)$ or $T_{b}(0)<T_{h}(0)$, while for $T_{b}(0)>T_{h}(0)$ apparently there is not violation. Moreover, in the future, for redshift $z=-1$, we have a divergence in the equations and therefore also a possible violation of the generalized second law, since in this case the system becomes unstable, implying arbitrary values of parameter $R$.

\subsection{Deformed Gauss-Bonnet dark energy}

\begin{figure}[htbp]
\begin{center}
\includegraphics[width=0.45\textwidth]{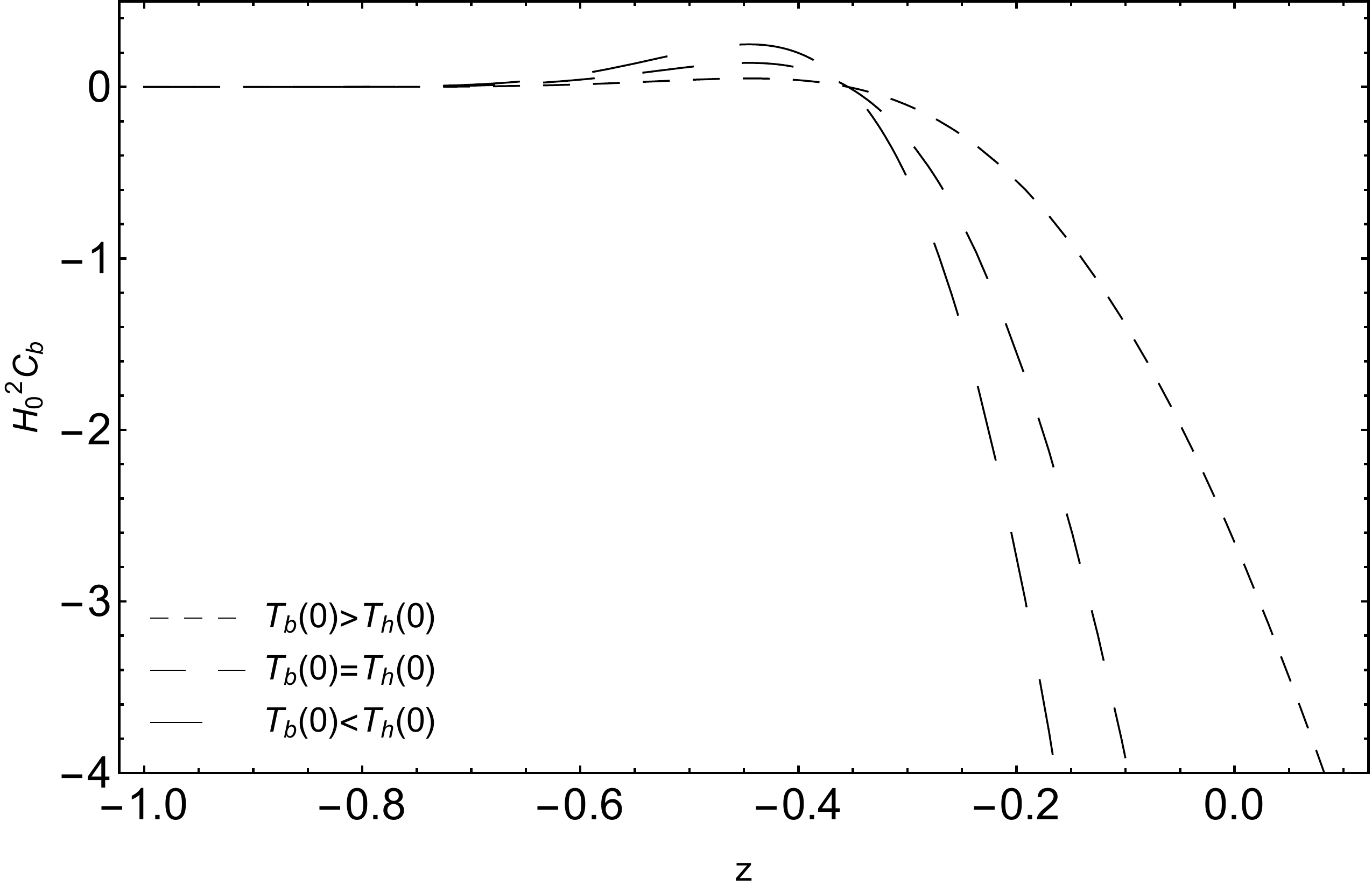}
\includegraphics[width=0.45\textwidth]{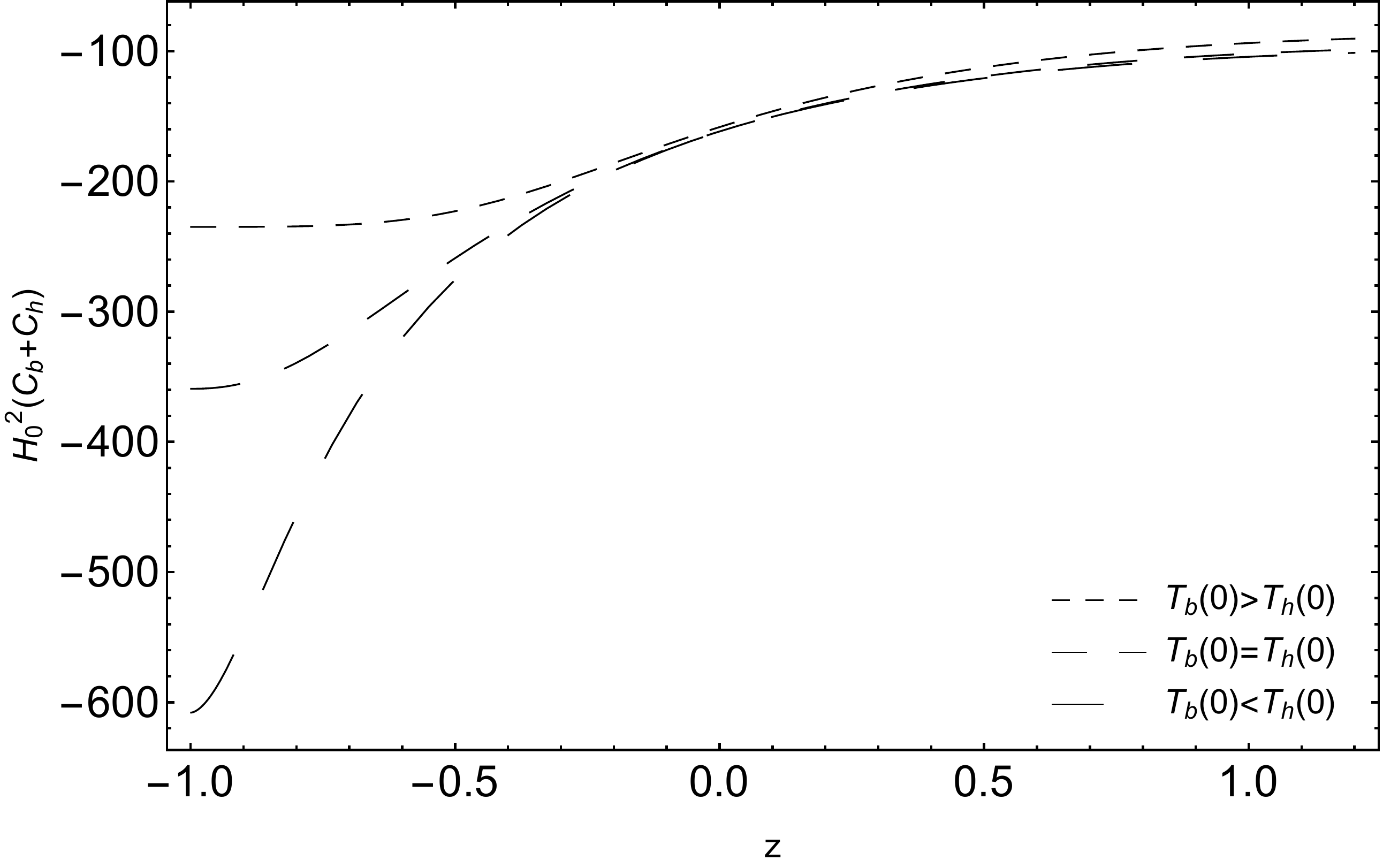}
\caption{\it{Deformed GB Model: we show the evolution of the heat capacity of the bulk (upper graph) and the behaviour of the sum of capacities of the bulk and the horizon (lower graph). Also, as in the case of GB model, from criteria $1$ and $2$ exposed in section \ref{Thermo}, we observe that the thermal equilibrium is possible as a transient phenomenon.}}
\label{FIG6}
\end{center}
\end{figure}

Following the analysis on thermal equilibrium and generalized second law in Ho\v{r}ava-Lifshitz cosmology we consider now the case of the deformed GB dark energy model. In FIG \ref{FIG6} (upper graph) we show the behaviour of the heat capacity of the bulk, whereas that in FIG \ref{FIG6} (lower graph) we depict the evolution of the sum of heat capacities of the bulk and the boundary (Hubble Horizon). As in the case of the GB model, we follow the criteria $1$ and $2$ for thermal equilibrium exposed in section \ref{Thermo}. We observe also a sign change of the heat capacity of the bulk in the future, at about $z\simeq -0.3$, which comes to take positive values later, being the sum of heat capacities, bulk plus boundary, always negative. Therefore we find that the thermal equilibrium is also a transient phenomenon in the deformed GB model, but, unlike GB model, it happens exclusively in the future.

In investigating the validity of the generalized second law we depict the behaviour of the parameter $R$ defined in Eq. \eqref{R}. We find different results in relation to the GB dark energy model, in particular, we observe that the generalized second law can be violated at higher redshifts, independently of the initial conditions $T_{b}(0)=T_{h}(0)$, $T_{b}(0)<T_{h}(0)$ and $T_{b}(0)>T_{h}(0)$. Also, we have a divergence at $z=-1$, which implies arbitrary values of $R$ and possible violations to the second law.

\begin{figure}[htbp]
\includegraphics[width=0.45\textwidth]{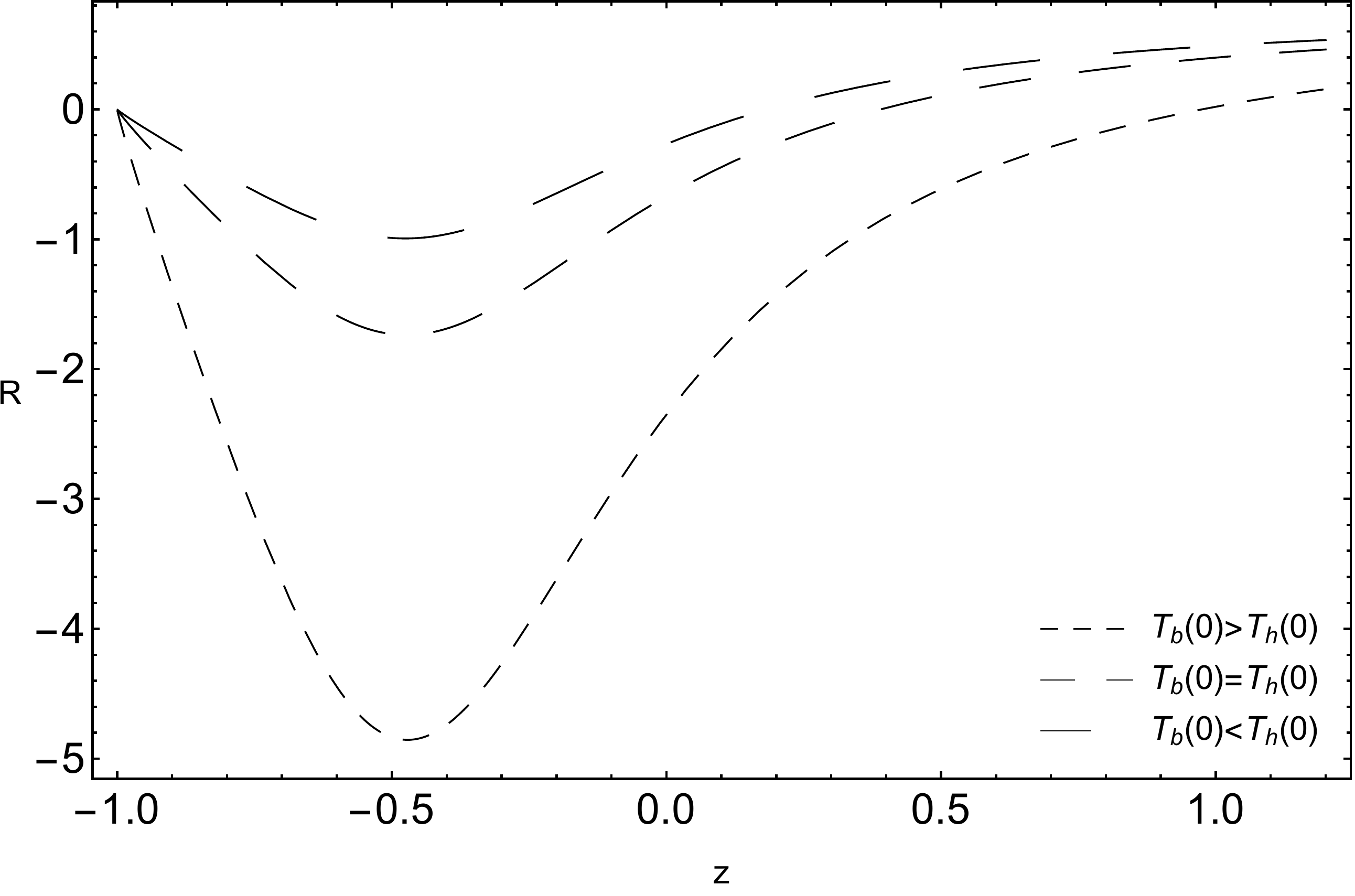}
\caption{\it{Deformed GB dark energy model: Evolution of parameter $R$ defined in Eq. \eqref{R}, which allows us to express the nonviolation of the generalized second law through the constraint $R\leq 0$. We observe there is a violation of the generalized second law at higher redshifts, independently of the initial conditions $T_{b}(0)=T_{h}(0)$, $T_{b}(0)<T_{h}(0)$ and $T_{b}(0)>T_{h}(0)$.}}
\label{FIG7}
\end{figure}

%%%%%%%%%%%%%%%%%%%%%%%%%%%%%%%%%%%%%%%%%%%%%%%%%%%%%%%%%%%%%%%%%%%%
\section{Concluding remarks}
%%%%%%%%%%%%%%%%%%%%%%%%%%%%%%%%%%%%%%%%%%%%%%%%%%%%%%%%%%%%%%%%%%%%
In investigating the bulk-boundary interaction and thermodynamics of the  Universe, the theory of Ho\v{r}ava-Lifshitz cosmology surges as an useful framework for this purpose. The $Q$ function describing the interchange of energy between the bulk (observable Universe) and the Hubble horizon (boundary) plays a fundamental role in determining the dynamics of the Universe and its thermal properties \cite{Cruz:2015lsz}. 

The energy density of the bulk is dominated by dark energy density and 
we have very interesting possibilities in choosing a good candidate for this dark energy component. Recently, in Ref. \cite{Granda:2013gka} it has been proposed a dark energy density based on the Gauss-Bonnet four-dimensional invariant and its modification, which can be physically interpreted as an energy density nonsaturating the Holographic principle 
\cite{Bekenstein:1973ur,tHooft:1993dmi,Bousso:1999xy,Cohen:1998zx,Susskind:1994vu}.  This Gauss-Bonnet (GB) energy density allows us to obtain the current values of dark energy density and dark energy EoS parameter compatible with the observational data \cite{Granda:2013gka}.

In the context of Ho\v{r}ava-Lifshitz cosmology, we investigated the bulk-boundary interaction for GB dark energy model and its deformed version. We discussed whether the thermal equilibrium is stable or not, once it is reached, and the validity of the generalized second law. In order to discuss whether the thermal equilibrium of the system bulk-boundary will be stable or not, once it is reached, it is necessary to study the evolution of the heat capacities of the bulk and the boundary. Whereas that the heat capacity of the Hubble horizon has a well defined sign which is always negative, the sign of the heat capacity of the bulk depends on the evolution and the sign of the $Q$ function \cite{Cruz:2015lsz}.

For both, GB dark energy and deformed GB dark energy model, we observed sign changes of the $Q$ function, and hence sign changes of the heat capacity of the bulk. Therefore, following the criteria exposed in section \ref{Thermo}, we can observe that for GB and deformed GB dark energy models, the thermal equilibrium can be reached only as a transient phenomenon, once that for both models $C_{b}$ takes positive values only during some time and the sum of capacities $C_{b}+C_{h}$ is negative. This happens independently of the set of initial conditions $T_{b}(0)>T_{h}(0)$, $T_{b}(0)=T_{h}(0)$ or $T_{b}(0)<T_{h}(0)$, for the temperature of the bulk and the boundary. Interestingly, in the case of deformed GB dark energy model, the $Q(z)$ function takes positives values at the present time, $z=0$, being this result important because positive values of $Q(0)$ are favoured from observational data (see e.g. \cite{Costa:2016tpb,Wang:2016lxa}), and hence, this could constitute an advantage in favour of the deformed model.  Here it is worth highlighting that in our results, for both models, GB and deformed GB dark energy models (FIGS. \ref{FIG1} and \ref{FIG3}), the values obtained by us for $Q(0)$ are nonzero at the present time, and we find $Q\rightarrow 0$ only for $z\rightarrow -1$. This means that the low energy limit can be recovered only in the future and not in the present.

On the other hand, the validity of the generalized second law is constrained by Eq. \eqref{R}, through the parameter $R$. In the case of GB dark energy model, we observed a violation to the second law at higher redshifts for the initial conditions $T_{b}(0)=T_{h}(0)$ or $T_{b}(0)<T_{h}(0)$, while for $T_{b}(0)>T_{h}(0)$ apparently there is not violation. In the case of deformed GB dark energy model, we found different results in relation to the GB dark energy model, in particular, we observed that the generalized second law can be violated at higher redshifts, independently of the initial conditions $T_{b}(0)=T_{h}(0)$, $T_{b}(0)<T_{h}(0)$ and $T_{b}(0)>T_{h}(0)$. However, for both models, we have a divergence at redshift $z=-1$, and therefore also a possible violation of the generalized second law at this redshift, since in this case the system becomes unstable, implying arbitrary values of parameter $R$.

\begin{acknowledgements}
S. Lepe acknowledges DI-PUCV $37.0/2018$ for financial support.
G. Otalora acknowledges DI-VRIEA for financial support through Proyecto Postdoctorado $2017$ VRIEA-PUCV.
\end{acknowledgements}

%%%%%%%%%%%%%%%%%%%%%%%%%%%%%%

\end{document}